\documentclass{emulateapj}
\usepackage{rotating}
\usepackage{color}
\def\cyan#1 {\textcolor{cyan}{#1}~}

\shorttitle{512-input Stream HPC Cross-Correlation}
\shortauthors{Kocz et al.}

\begin{document}

\title{Digital Signal Processing using Stream High Performance Computing: A 512-input Broadband Correlator for Radio Astronomy}

\author{J. Kocz$^{1,9\dagger}$, L.J Greenhill$^1$, B.R. Barsdell$^1$, D. Price$^1$, G. Bernardi$^{2,3}$, S. Bourke$^4$, M.A. Clark$^{1,4,5}$, J. Craig$^6$, M. Dexter$^7$, J. Dowell$^6$, T. Eftekhari$^6$, S. Ellingson$^8$, G. Hallinan$^4$, J. Hartman$^9$, A. Jameson$^{10}$, D. MacMahon$^7$, G. Taylor$^6$, F. Schinzel$^6$, D. Werthimer$^{11}$}

\address{
$^1$ Harvard-Smithsonian Center for Astrophysics, 60 Garden Street,
Cambridge, Massachusetts, 02138, USA\\
$^2$ SKA SA, 3rd Floor, The Park, Park Road, Pinelands, 7405, South Africa\\
$^3$ Department of Physics and Electronics, Rhodes University, PO Box 94, Grahamstown, 6140, South Africa\\
$^4$ California Institute of Technology, 1200 E. California Blvd., Pasadena 91125, USA\\
$^5$ NVIDIA Corporation, 2701 San Tomas Expressway, Santa Clara, CA 95050, USA\\
$^6$ Department of Physics and Astronomy, University of New Mexico, Albuquerque, NM 87131, USA\\
$^7$ Radio Astronomy Laboratory, University of California, Berkeley, CA 94720, USA \\
$^8$ Bradley Department of Electrical and Computer Engineering, Virginia Tech, Blacksburg, VA 24061, USA \\
$^9$ Jet Propulsion Laboratory, California Institute of Technology, Pasadena, CA 91109 USA\\
$^{10}$ Centre for Astrophysics and Supercomputing, Swinburne University of Technology, 1 Alfred Street, Hawthorn, Victoria, 3122, Australia \\
$^{11}$Space Sciences Lab, University of California, Berkeley, CA 94720, USA 
}

\begin{abstract}
A ``large-N'' correlator that makes use of Field Programmable Gate Arrays and Graphics Processing Units has been deployed as the digital signal processing system for the Long Wavelength Array station at Owens Valley Radio Observatory (LWA-OV), to enable the Large Aperture Experiment to Detect the Dark Ages (LEDA).  The system samples a $\sim100$\,MHz baseband and processes signals from 512 antennas (256 dual polarization) over a $\sim 58$\,MHz instantaneous sub-band, achieving 16.8\,Tops\,s$^{-1}$ and 0.236 Tbit\,s$^{-1}$ throughput  in a 9\,kW envelope and single rack footprint. The output data rate is 260~MB\,s$^{-1}$ for 9 second time averaging of cross-power and 1 second averaging of total-power data. At deployment, the LWA-OV correlator was the largest in production in terms of $N$ and is the third largest in terms of complex multiply accumulations,  after the Very Large Array and Atacama Large Millimeter Array.    The correlator's comparatively fast development time and low cost establish a practical foundation for the scalability of a modular, heterogeneous, computing architecture.

\end{abstract}


\section{Introduction}

The Large Aperture Experiment to Detect the Dark Ages \citep[LEDA;][]{greenhill2011} is designed for study of the sky-averaged absorption spectrum of the HI 21\,cm line from the intergalactic medium at z$\sim$20 seen against the Cosmic Microwave Background.  The scientific goal is characterization of the thermal history of the universe through to the end of the cosmological Dark Age and onset of X-ray heating of the intergalactic medium by the accumulating end products of stellar evolution.  In principle, a lone dipole can  measure the sky-averaged spectrum, but LEDA has embedded antennas outfitted for radiometry in a large-N interferometric array that may be used to directly derive particular critical calibrations  {\it in situ}   \citep[i.e., sky models, broadband polarized antenna gain patterns, and ionospheric refractive offsets and (de)focusing;][]{bernardi2014}.

In targeting a 30-88\,MHz passband, LEDA presents a signal processing challenge in correlation because it requires an array with a large number of interferometric elements (a.k.a. ``large N''). 

Large N enables reconstruction of the predominant diffuse brightness distribution of the low-frequency sky and assembly of a sky model.  Coupled with moderate baselines that suppress confusion noise (up to on the order of 1 km), large-N provides sensitivity to point source calibrators for (i) estimation of antenna patterns from direction dependent gain measurements and (ii) 2-D modeling of the ionosphere from time-variable refractive offsets \cite{mitchell2008}. The experiment is deployed to two stations of the Long Wavelength Array \citep[LWA;][]{taylor2012, ellingson2013}: the LWA1 site in New Mexico and the Caltech Owens Valley Radio Observatory site in California (LWA-OV). A 64 input system processing 57.6\,MHz of instantaneous bandwidth is deployed at LWA1, and at LWA-OV, LEDA has deployed a full correlation system that provides cross and total power spectra for all 512 inputs with an instantaneous bandwidth of 57.552\,MHz.

The LEDA correlator provides an early demonstration of a large scale hybrid architecture that combines Field Programmable Gate Arrays (FPGA) for $\cal{O}$(N) calculation of spectra for each antenna at the Nyquist rate and Graphics Processing Units (GPUs) for $\cal{O}$(N$^2$) cross multiplication, in real time.   Description of the parallelized and scalable FX architecture is described in \citet[][hereafter K14]{kocz2014}, as well as in the field demonstration of a 32-input system. First light for the $N=512$ system was obtained on 29 July 2013. The following sections present implementation of the 512-input LEDA design, including hardware and firmware/software (\S\ref{sec:ware}), performance characteristics (\S\ref{sec:performance}), monitor and control (\S\ref{sec:gui}), and demonstration of first light (\S\ref{sec:first light}). 

\section{Specifications, Hardware \& Performance}
\label{sec:ware}

\begin{figure}
\begin{center}
\includegraphics[width=9cm]{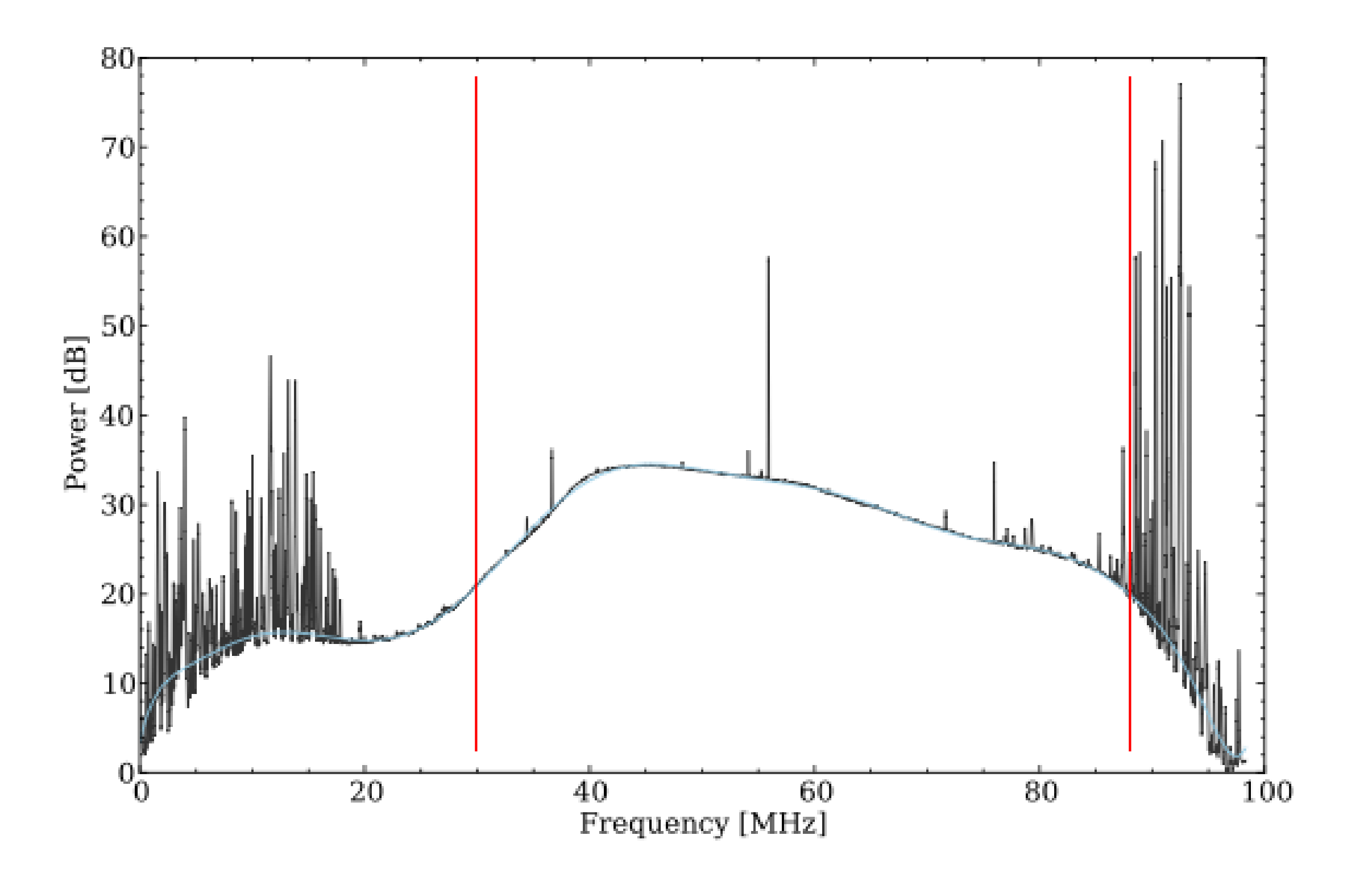}
\caption{Median power spectrum over 36-hours from 30-Sep to 1-Oct 2013. The full spectrum is shown in black, with a $13^{th}$ order polynomial bandpass fit overlaid in blue.  Red lines denote the correlated passband.  The roll off and shoulder below 40\,MHz reflect the response of an HF-band interference suppression filter in the LWA analog path.  In the FM radio band, interference can peak at 60\,dB above the sky even with notch filtering (50\,dB shown here).  The FM spectrum is aliased at a clock frequency of 98.304\,MHz  \label{fig:rfi}}
\end{center}
\end{figure}

Correlator signal processing specifications (Table\,1) stem principally from the primary science driver (detection of the HI absorption trough), array layout, calibration requirements, and local interference conditions at LWA-OV (Figure\,\ref{fig:rfi}).  A 212\,m diameter core of 251 dual polarization antennas is supplemented by an arc of 5 additional (outrigger) antennas with a radius 265\,m.  The longest baseline is 504\,m, and the longest core-outrigger baseline is 373\,m.\footnote{The LWA1 configuration comprises a $100\times110$\,m core and five outriggers distributed along an offset Reuleaux triangle with baselines up to $\sim 560$\,m.} The correlator is non-tracking (i.e., topocentric) because the LWA dipoles are fixed, and the breadth of the field of view is nearly a full hemisphere ($\sim 140^\circ$ across at -10\,dB).  Nearly a full 100\,MHz baseband is digitized into 4096 channels, affording 24.0\,kHz channel spacing that benefits detection and excision of interference.   The fine channelization also limits decorrelation on the longest core-outrigger baselines to $<1\%$.  (A 47\,kHz channel and 500\, m baseline would suffer 1\% loss. A 300\,m baseline suffers the same loss in a 78\,kHz channel.)  Computing resources permit stream processing of 57.552\,MHz (2398 channels), several times broader than the anticipated of approximately 10\,MHz wide Dark Ages signal for a range of cosmological models.  Decorrelation due to time integration is also $<1\%$ on the longest core-outrigger baseline for a 9 second integration time.  This is chosen to be a multiple of 3 seconds because of the switching cycle between sky and calibration paths within the array (see Section\,\ref{sec:fast}).

In order to serve the number of inputs and observing bandwidth, the $\cal{O}$(N) and $\cal{O}$(N$^2$) stages of the correlator must execute 1.8 and 15.0\,Tops\,s$^{-1}$  on a 0.236~Tbit\,s$^{-1}$ stream with an initial sustained output data rate of 260\,MB\,s$^{-1}$ (see K14 for scaling relations).  The remote location and size of the shelter available for the correlator constrained power dissipation to a ceiling of 12\,kW and size to a maximum footprint of one 42U rack.  The correlator computing capacity is comparable to a third of that of the Very Large Array (16.41 and 23.7624\,Tops\,s$^{-1}$) of the National Radio Astronomy Observatory when correlating 54 inputs of 8\,GHz each.  The motivation for selecting each hardware element is discussed in \S\ref{ss:digi} - \S\ref{ss:gpu}.  Table\,\ref{table:hardware} summarizes the configuration.

\begin{table}
\begin{center}
\caption{Signal Processing Requirements} 
\begin{tabular}[t]{l@{\extracolsep{-0.03in}}c@{\extracolsep{-0.0in}}c@{\extracolsep{+0.01in}}rl}
\hline
Specification & ~Requirement& ~Actual & Driver \\
\hline
\hline
Inputs           & 512 & 512$^{\rm (a)}$ & Calibration\\
Baseband per polarization & 1 & 1 & Science \\
Baseband sampled (MHz)& 0-98   & 98.304& Science, RFI\\
Correlated bandwidth (MHz) & 50 & 57.552 & Science$^{\rm (b)}$\\
Channelization (kHz) &  30 & 24.0 & Decorrelation, RFI\\ 
Dump time--sec (x/c) & 24 & 9& Decorrelation$^{\rm (c)}$\\
\hbox to 0.83in{  }(a/c) & 1 & 1 & Calibration$^{\rm (d)}$\\
Output rate (Gb\,s$^{-1}$)  & 1  & 2.24 & Cost$^{\rm (e)}$\\
Power (kW) & 12 & 9 & Remote location$^{\rm (f)}$\\
Volume (RU) & 42 & 36 & Remote location\\
\hline
\end{tabular}
\end{center}
\hspace{0.055in}$^{\rm(a)}$256 antennas, dual polarization. Products: XX, YY, XY, YX.\\
\hbox to 0.05in{}$^{\rm(b)}$Predicted absorption trough width is $\sim 10$\,MHz (FWHM) \\
\hbox to 0.23in{}at 60\,MHz \citep[e.g.,][]{pritchard2010}, motivating study \\
\hbox to 0.23in{}of the $\sim$30-88\,MHz band, between HF and FM broadcast.\\
\hbox to 0.05in{}$^{\rm(c)}$Fringe rate decorrelation in time averaging $<$0.8\% for the\\
\hbox to 0.23in{}longest core-outrigger baseline at LWA-OV, 373\,m, at 88\,MHz.\\
\hbox to 0.23in{}The current minimum dump time is 0.33 sec (60.5\,Gb\,s$^{-1}$).\\
\hbox to 0.05in{}$^{\rm(d)}$Three state calibration cycle is 3 sec., triggered by a 1\,PPS.\\
\hbox to 0.05in{}$^{\rm(e)}$Higher rates require increased on-site storage.\\
\hbox to 0.05in{}$^{\rm(f)}$Under computational load. Three circuits: 208VAC 30A each.\\ 
\label{table:reqs}
\end{table}

\begin{table}
\begin{center}
\label{table:hardware}
\caption{LEDA-512 Equipment list}
\begin{tabular}[t]{c|c|c}
\hline
Subsystem         &  Hardware           & Notes\\
\hline
\hline
F stage           & 16x ROACH2          & 2x16 input ADCs each       \\
                  &                             & 8192-pt PFB \\
                  &   (1U each)          & 10\,GbE `packetizer'\\
                  &                            & 196.608~MHz clock input\\
\hline
Interconnect      &   Switch            & 12/48-port 40/10\,GbE \\
                  & (1U)                & Mellanox SX1024 QSFP/SFP+\\
\hline
X stage           & 11x Server          & Dual-E2670 CPUs\\
                  & (1U each)           & Dual NVIDIA K20X GPUs\\
                  &                     & 40\,GbE port\\ 
                  &                     & 128~GB RAM\\
                  &                     & Dual 2~TB HDD\\
\hline
Data out    & Switch    & 24-port 1\,GbE  \\
                  & (1U)        & 2-port 10\,GbE\\
                   &               &HP 2920-24G\\
\hline
Clock             & Valon 5008          & 196.608~MHz\\
                  & Synthesizer         & 32-way amplifier\\
                  & (2U)                & and splitter\\
\hline
PPS               & 16-way PPS          & Linked to Trimble\\
                  & distributor         & Thunderbolt E GPS\\
                  & (1U)                & Disciplined Clock\\
\hline

Admin             & Switch      & 48-port 1\,GbE\\
                  & (1U)                & HP 2920-48G\\
                  & Headnode            & Intel Xeon E3-1230v2\\
                  & (1U)                & 6x 1\,GbE ports\\

\hline
\end{tabular}
\vbox to 0.1in{}
\end{center}
\end{table}

\begin{figure}
\begin{center}
\includegraphics[width=5cm]{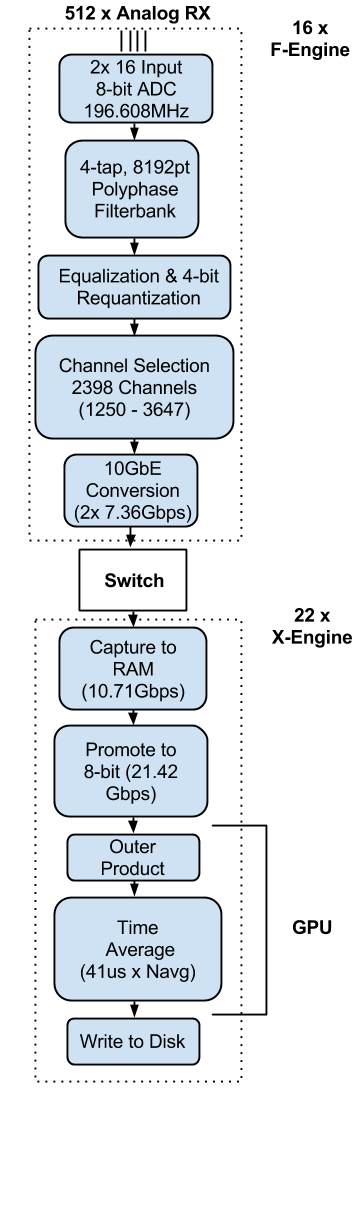}
\caption{Correlator primary data flow elements and corresponding data rates.\label{fig:l512_dataflow}.  Data transport is unidirectional, from the F-engines (FPGA-layer), which  transform the time domain data stream for each antenna, to the X-engines (GPU layer), which cross multiply and time average these signals, antenna pair by antenna pair. The transpose operation between stages (i.e., corner turn) is executed via network routing and a fine data re-ordering step in the CPU layer.  (No FPGA resources are used.)  This implementation enabled ready scaling to $N=512$ and can be extended by at least an order magnitude in $N$ (K14).}
\end{center}
\end{figure}

\subsection{Signal digitization}
\label{ss:digi}
The signals are digitized using an 8-bit, 16-input ADC (ADC16x250)\footnote{General specifications of the ADC board used can be found at https://casper.berkeley.edu/wiki/ADC16x250-8\_RJ45\_rev\_1}. Four RJ45 jacks ($4\times4$ differential pairs total) provide high density differential  input from the LWA analog path.  The Hittite HMCAD1511 used on this board was chosen for its high dynamic range, small footprint, quad-channel mode, and differential input.   In addition, this ADC affords flexibility regarding the required analog path gain because it samples 12 bits total (exposing only 8) and features selectable digital gains, though the effective number of bits sampled is reduced for large gain settings.  Bench measurements of the ADC16x250 board show cross-talk among inputs sharing the same chip to be below $\sim -25$\,dB between 30 and 88\,MHz for the digital gain settings adopted by LEDA.\footnote{https://casper.berkeley.edu/wiki/ADC16\_cross\_corr\_tests}  Cross-talk among channels on different chips is  $\sim -50$\,dB.  Preliminary tests suggest that rejection 5-10 dB better may be achieved for particular RJ45 conductor pairs (12, 78) and paths within the ADC IC (2,4) if the others receive no signal.

A 2\,dBm, 196.608\,MHz signal synthesized by a Valon 5008 synthesizer and locked to the 10\,MHz output of a Trimble Thunderbolt E GPS disciplined clock, to prevent frequency drift, clocks the ADC. A synchronization pulse from a GPS locked clock is sent to each FPGA which determines the start time of an observation. A 16-way distributor connected to the GPS locked clock provides the pulse. The average delay through the splitter is $\sim$14\,ns. As each ADC is clocked from the same source, sampling and FPGA processing is synchronous.  Because time tags downstream are referenced to the observation start, even asynchronous network and GPU computing elements are effectively synchronous too. Manual delay calibration is required for each of the ADC16-250-8 cards on power up. This covers time sample offsets among the chips on each card (up to +/- one clock cycle) as well as between the cards.

\subsection{FPGA Layer}

A Xilinx Virtex-6 SX475T FPGA is used for the polyphase filterbank (PFB), requantization, channel selection and packetization of the data. The FGPAs are mounted on a ROACH2 board developed by CASPER\footnote{https://casper.berkeley.edu/wiki/ROACH-2\_Revision\_2}\cite{parsons2006}. This allows direct connection of the ADC boards using a ZDOK connector (two per ROACH2).  Each ROACH2 board has a daughter board with  four 10\,GbE SFP+ ports. Two of these are used to transmit 7.37\,Gbit\,s$^{-1}$ each. Packets are sent alternately from the outputs so as not to saturate the 10\,GbE link. 

The theoretical peak performace of a Virtex-6 SX475T is 1.2\,Top\,s$^{-1}$, assuming that all 2016 DSP48E slices are in use, and a clock rate of 600\,MHz can be achieved. In practice, this clock rate and usage are rarely achieved. Assuming a realistic maximum clock rate for this design of 250\,MHz, then based on implemented  DSP48E usage (47\%) and clock (196.608\,MHz), the FPGA runs at 36\% of maximum performance. The primary limitation of the design in its current form is the FPGA block RAM (BRAM), with usage currently at 82\%. This is primarily due to coefficient storage requirements for the PFB, limiting the PFB to 4 taps. Data packets are also buffered in BRAM prior to transmission. Should additional buffering be required, it is likely this would need to be moved to the attached SRAM memory.

A 4-tap, 8192-point PFB transforms the data into the frequency domain with resolution of 24.0\,kHz (Figure\,\ref{fig:l512_dataflow}). The PFB is implemented on the FPGA as an FIR filter followed by an FFT \cite{parsons2009}. The PFB has 18-bit coefficients and uses a Hamming window to sharpen the filter response. Interference from broadcast can be many tens of dB stronger than the sky brightness in some  lower VHF sub-bands, and this creates specific demands for the FFT stage. The channel isolation of the 4 tap PFB exceeds 70\,dB for adjacent channels  (opposed to a non-windowed FFT, where the first sidelobe is at 13.6\,dB), thereby better restricting the impact of radio frequency interference (RFI) to narrow channel ranges. The 18+18-bit complex coefficients of the FFT limits risk of overflow, with further insurance provided by a right shift (divide by 2) at each internal multiplication stage.  After the PFB, the data are requantized from 18+18 bit depth to 4+4 bits, reducing the output data rate by $4.5\times$ without a substantive increase in quantization noise \citep[e.g.,][]{tms01, thompson2007}. To ensure effective use of the four bits, prior to requantization, an average digital gain factor is applied across the band. More sophisticated schemes in which gains are estimated and applied channel by channel could also be implemented.  The cost of using a single factor is marginal because high quantization noise is limited to the band edges, where the RF response rolls off considerably (Figure\,\ref{fig:rfi}).

The final stage of FPGA processing formats data into UDP packets. A group of 2398 channels (corresponding to the frequency range 30-88\,MHz) are selected for this process. The purpose of this is twofold. Firstly, the data below 30\,MHz and above 88\,MHz are corrupted by HF and FM radio broadcasting. Secondly, it reduces the load on the network, and the computational requirements of the cross-multiplication stage of the correlator.  Any 58\,MHz within the sampled 0-98\,MHz baseband can be selected via run-time parameters and without firmware recompilation. The UDP data segment contains a 128-bit header specifying the order sequence and 6976 bytes of data, corresponding to eight, 8-bit samples for 109 channels. The use of a packetized architecture \citep{parsons2008} also allows for the corner turn (reordering of the data so that the inputs are grouped by frequency instead of time), to be completed without additional processing overhead (K14).

\subsection{Network Layer}
\label{ss:network}

The FPGA (F-engine) and CPU/GPU (X-engine) elements of the system are networked via a Mellanox SX1024 switch. With 12 40\,GbE ports and 48 10\,GbE ports, this switch supports dual 10\,GbE connections to each ROACH2, and a single 40\,GbE connection to each server. Using 40\,GbE in the servers simplifies the network configuration (vs N$\times$10\,GbE), as each server only has one capture port on the network subnet, and gives room for expansion should servers be required to process larger bandwidths. 

\subsection{CPU Layer}
\label{ss:cpu}
The two primary operations executed in the CPU layer are data capture from the network and unpacking the data (from four to eight bit) to stage it for transfer to the GPUs.
Data are captured to RAM using the \texttt{PSRDADA}\footnote{http://psrdada.sourceforge.net/} software package. \texttt{PSRDADA} monitors the input network port using a busy-wait loop, copying packets to ring buffers in memory as they are received. Using this software capture rates per CPU core up to 15\,Gb\,s$^{-1}$ have been demonstrated. For the LEDA-512 correlator, each 40\,GbE port captures two streams of 10.71\,Gb\,s$^{-1}$, with each stream bound to a separate CPU core. Each CPU process (e.g., data capture, unpack, transfer to GPU) is bound to a separate core. This is required to achieve consistently lossless data transfer at rates $\gg 10$\,Gb\,s$^{-1}$.

\begin{figure}
\begin{center}
\includegraphics[width=9cm]{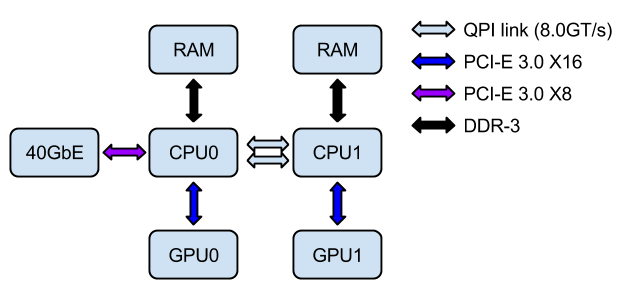}
\caption{Data flow between the two e5-2670 CPUs in each X-engine node and the network card, host memory, and GPUs. \label{fig:cpu}}
\end{center}
\end{figure}

Hardware resources serving each data stream (i.e., CPU, GPU, RAM, and buses) are kept as disjoint as possible, so as to minimize bottlenecks (Figure \ref{fig:cpu}). For instance, the \texttt{PSRDADA} ring buffers that serve each of the two GPUs in a server are hosted in RAM for different CPUs. The only high bandwidth communication between CPUs involves the single 40\,GbE interface (hosted on CPU0).  Data for CPU1 must be transferred over the motherboard QPI bus during initial capture. However, no issue with lost packets has been observed at the required capture rates. Under typical conditions, running tallies show packet loss of  $\sim 1.5$ in every $10^{9}$ packets. Each packet contains a 16\,B header and 6976\,B of data. Since the amount of data lost due to dropped packets is negligibly small, no correction for packet loss is applied.  

Unpacking involves promotion of data from 4+4 to 8+8 bit complex representation and re-ordering.  Promotion from 8+8 to the 32\,bit floating point representation required by GPU arithmetic  units is executed on the fly by GPU texture hardware within the cross-correlation pipeline \citep{clark2012}.  Absence of such ``free'' promotion from 4 to 32 bits, and consideration of PCIe bus transfer times motivate the use of CPU resources for the first ``half,'' thereby enabling GPU resources to be entirely focused on $\cal{O}$(N$^2$) cross-multiplication.  After promotion, data are re-ordered to complete the corner turn operation begun in network hardware without drawing upon general computational resources (K14).  A single CPU core can unpack $\sim 16$\,Gb\,s$^{-1}$ without loss. Multiple CPU cores can be assigned to achieve higher rates.   

\subsection{GPU Layer}
\label{ss:gpu}

X-engine processing using the \texttt{xGPU}\footnote{https://github.com/GPU-correlators/xGPU} library \citep{clark2012} generates visibilities with 9 second integration times.  Data are read  in groups of 8000 spectra, cross multiplied and averaged, resulting in a base time resolution of 0.33 seconds.  Averaging of 27 groups yields the default 9 second cadence (Table\,\ref{table:reqs}). Data are copied back to \texttt{PSRDADA} ring buffers in RAM and written in separate streams (one per GPU) to a remote mass storage server at 2.24\,Gb\,s$^{-1}$ ($\sim 1$\,TB\,hr$^{-1}$).  Each K20X GPU (GK110 architecture) achieves $\sim 3$\,TFlop\,s$^{-1}$ (sustained), corresponding to 56\% single precision resource utilization, for a clock rate of 1.002\,GHz, and a data throughput of 21.4 \,Gb\,s$^{-1}$  per GPU (42.8\,Gb\,s$^{-1}$ per server).  

Selection of the Tesla series Kepler GPU was driven by the need for a high density of single precision resources, aggregation of computation on a single die, and low power consumption.  For the Kepler microarchitectures available during development, the GTX Titan, GTX~690, K10, and K20X have relevant compute densities.  The former two are excluded because the correlator operation within 42U requires a density of two GPUs per 1U (see Section\,\ref{sec:ware}), and the server chassis cannot accept consumer grade cards due to size, ventilation, and wiring constraints. The middle two GPUs are excluded because as dual-die cards, they require at least 8 CPU cores for management of data streaming from NIC to GPU, for the current optimized code (K14).  These overriding considerations drove adoption of the K20X.  

\subsection{High Cadence Output of Total Power Data}
\label{sec:fast}

While visibility data is integrated over 9 seconds, auto-correlation spectra for the (10) outrigger antennas that are to be calibrated for measurement of the 21cm signature at $z\sim 20$ are integrated for only 1 second.   This differential integration is executed in the GPU layer via selective zero-ing and dumping of cells in the correlation matrix accumulator.  The more rapid cadence caters to the monitor and control system adopted by LEDA, where a 1\,PPS signal is used to trigger switching in a  three state calibration cycle employed to estimate calibrated sky temperature \citep[e.g.,][]{bowman2010}.   The aggregate data rate for 1 second averaging of the 10 outrigger antennas is small compared to the correlator output ($\sim 0.8$\,Mb\,s$^{-1}$), and to ensure the main correlator pipeline is not disrupted, the GPU is bypassed and the corresponding mode writes 4-bit requantized F-engine output for selected outrigger antennas separately to RAM. This one second integration time can be arbitrarily reduced to any value up to the Nyquist rate (1/24.0 kHz or 41.67 $\mu$s), limited only by disk write speed. 

\subsection{Data transport \& Network Management}
\label{ss:dataTransportation}

\begin{figure*}
\begin{center}
\includegraphics[width=6in]{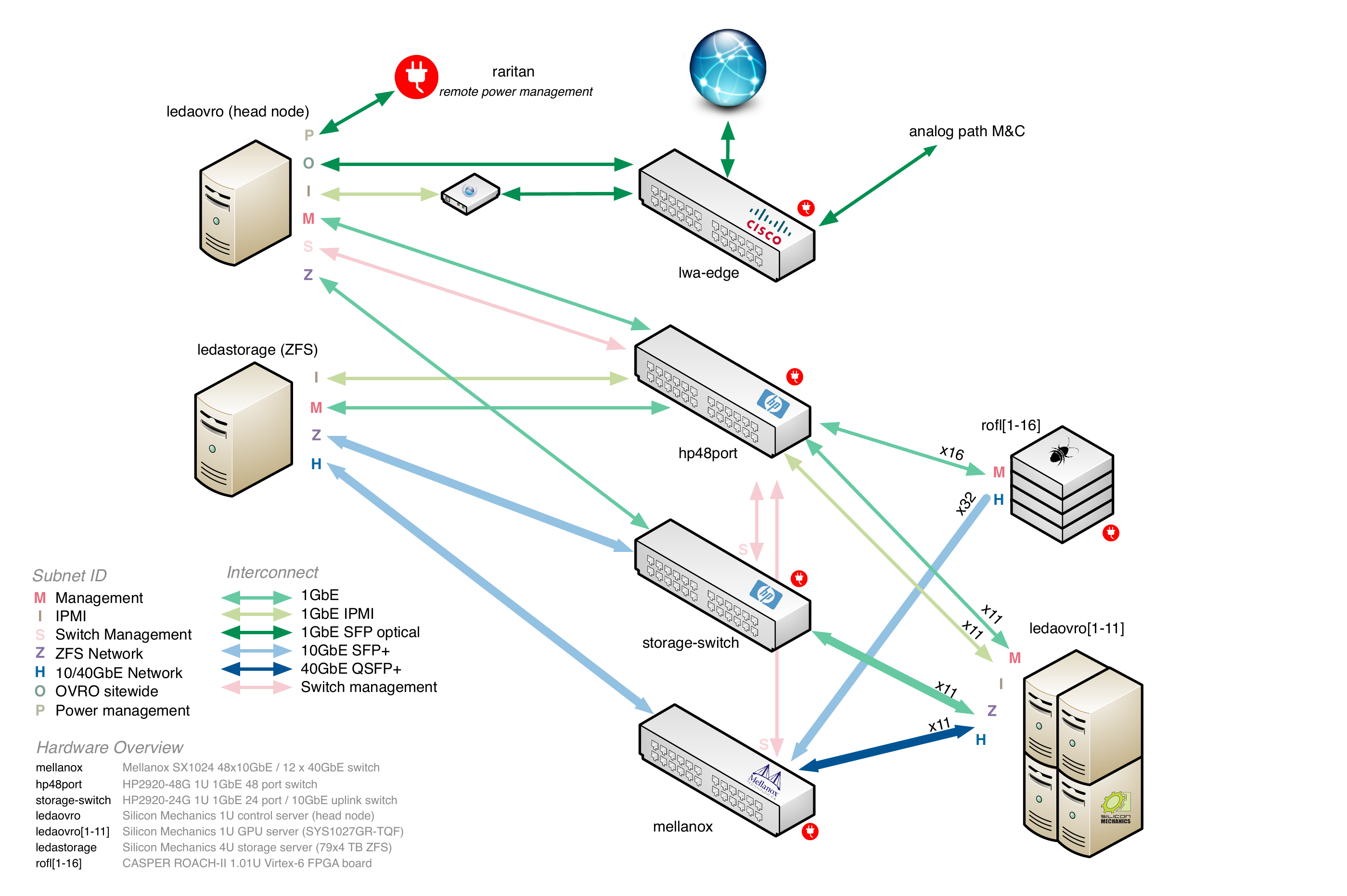}
\caption{LEDA correlator network. Color codes the type of interconnect among nodes and switches.  Upper case letters code the purpose of traffic. Critical systems controlled by an Ethernet enabled switched power distribution unit  are indicated by a red symbol.}
\label{fig:network}
\end{center}
\end{figure*}

The network architecture emphasizes unidirectional high throughput transfer of data from  the FPGA nodes to a network switch and thence to the GPU servers (Figure\,\ref{fig:network}).  The network uses Ethernet standard rather than Infiniband, due both to the Open nature of the standard, and the use of UDP packets.  A 10/40\,GbE Mellanox SX1024 is used to enable distribution of 10\,GbE traffic across 40\,GbE links. The SX1024 is run with flow control enabled on the 40\,GbE ports, however no other configuration changes were necessary for the required data rate (236\,Gb$s^{-1}$. The 40\,GbE cards in the servers are configured with the Mellanox ethernet drivers. Some changes were made to the operating system parameters, outlined in Table\,3,   to ensure the smooth capture of data and flow through the \texttt{PSRDADA} pipeline. 

\begin{table}
\begin{center}
\caption{Network/Memory OS Configuration}
\begin{tabular}[t]{cc}
\hline
Parameter    & Value\\
\hline
\hline
\\
kernel.shmmax & 68719476736\\
kernel.shmall & 4294967296\\
net.core.netdev\_max\_backlog & 250000\\
net.core.wmem\_max & 536870912\\
net.core.rmem\_max & 536870912\\
net.core.rmem\_default & 16777216\\
net.core.wmem\_default & 16777216\\
net.core.optmem\_max & 16777216\\
net.ipv4.tcp\_mem & 16777216 16777216 16777216\\
net.ipv4.tcp\_rmem & 4096 87380 16777216\\
net.ipv4.tcp\_wmem & 4096 87380 16777216\\
net.ipv4.tcp\_timestamps & 0\\
net.ipv4.tcp\_sack & 0\\
net.ipv4.tcp\_low\_latency & 1\\
\hline
\end{tabular}
\end{center}
\label{tab:net}
\end{table}

Two secondary requirements driving network architecture are that (i) visibilities are streamed from each GPU without aggregation to form monolithic cross-power spectra prior to storage, and (ii) following power or other glitches, control over the FPGA/GPU cluster is failsafe.  The former requirement traces from the design of the intended post-correlation data processing system, which is parallelized over frequency \citep[e.g.,][]{edgar10}.  In response, the correlator writes the multiple streams to a lightweight 246\,TB  ZFS RAID storage server with 5\,Gb\,s$^{-1}$ write capacity (sustained).  The latter requirement stems from the  necessity for unattended failsafe operation at a remote location.   Only two elements need to be functional to enable recovery of the correlator:  the external network switch and the Trendnet TFC-1000MGA RJ-45 to SFP 1\,GbE adapter that services the IPMI link to the head node.   All other IPMI and switched power systems (switches, nodes, and Raritan PX2-2496 Ethernet controlled power strip) can be booted and managed from the head node.  As they are critical components, all switches within the correlator  are powered from the Raritan strip so that they may be hard cycled if necessary.

\section{Power Dissipation}
\label{sec:performance}

\begin{figure}
\begin{center}
\includegraphics[width=3in]{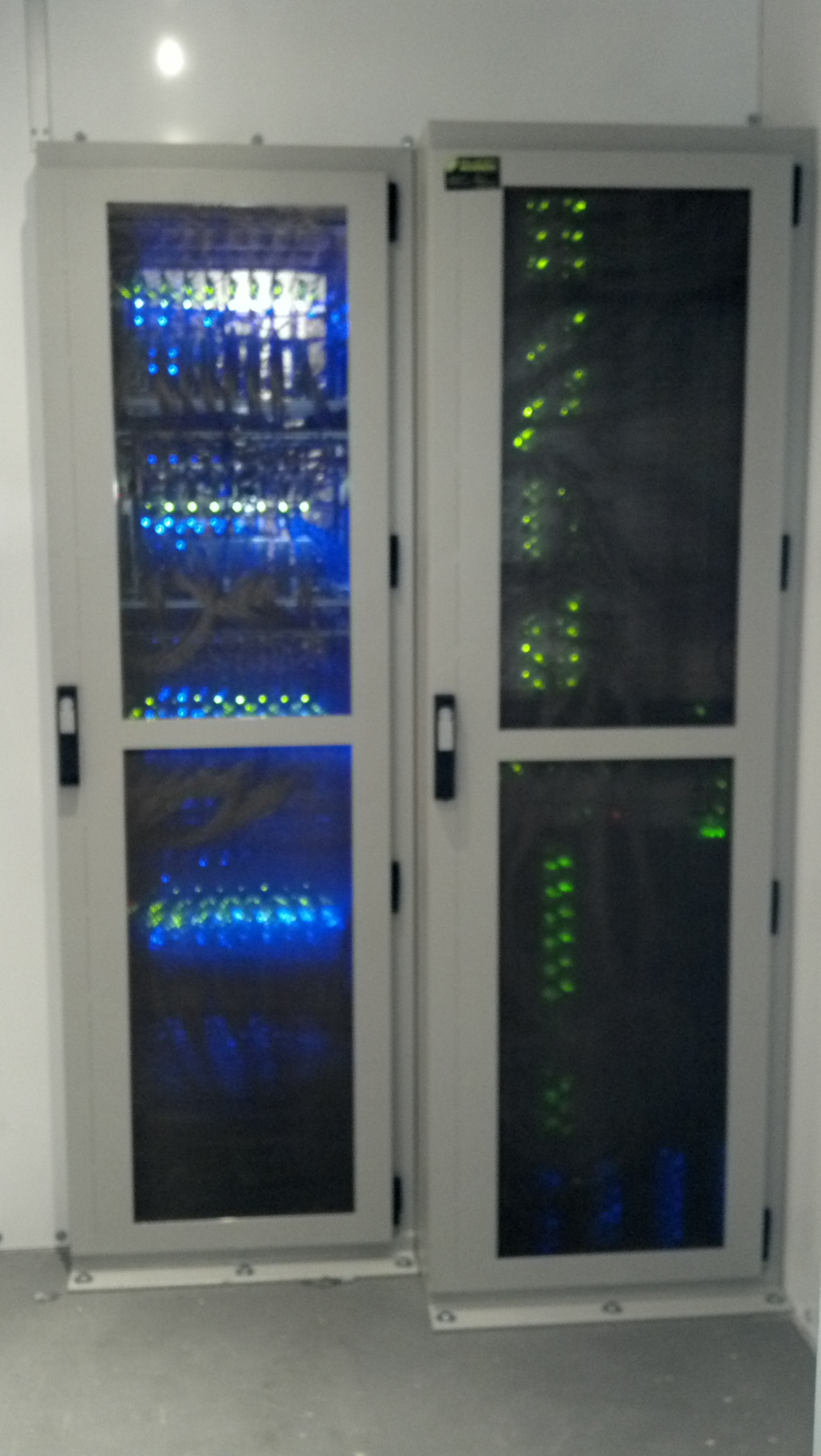}
\caption{Front view of LEDA correlator racking (right).  Large-format EMI air filters adopted to enable most readily the front-back airflow required by the digital hardware comprise the door.  Screening at 40\,MHz is $\sim50$\,dB.  The lefthand rack contains high-gain analog electronics and is independently shielded against RF interference.    RF-tight pass throughs in the common wall convey signals from the analog electronics to the correlator digital samplers.  The setup is conceptually similar to that adopted for LWA1 \citep[]{taylor2012, ellingson2013} though more compact and capable of supporting greater power and computational density.\label{fig:l512_rack}}
\end{center}
\end{figure}

High power density and heat dissipation within the constrained 42U footprint of the correlator are primary practical challenges.  An air cooling model was adopted to accomplish this.   For a 9\,kW budget, of which 7.4\,kW is consumed by the GPU servers (670\,W per 5.23\,MHz), constraining the  temperature fluctuation in the rack to $< 5^\circ$\,C requires $\sim 3200$\,CFM.  Off-the-shelf computing components  used in the correlator (e.g., servers) are engineered for front-to-back airflow.   The adopted rack design balances this need with a requirement for RF shielding of the digital systems ($>40$\,dB up to 20\,GHz) driven by proximity of high-gain analog systems as well as the antenna array outside the shelter (Figure \ref{fig:l512_rack}).   Large-format electromagnetic interference $1/8''$ Al honeycomb air filters ($17''\times36''$; Spira-EMI) are mounted in cutouts in the front and rear doors above and below waist height.  An additional screen covers a cutout in the rear portion of the rack roof to vent heat not advected in the flow through the rear filters.  Transmitter-receiver testing at 40\,MHz suggest attenuation of $\sim 50$\,dB in the lower VHF band.  (Screening of the machine room contributes an additional $\sim 60$\,dB.)   Air is drawn through the front door of the rack by the OEM fan arrays built into the servers, switches, and other hardware.  A cold reservoir is assured because a partition divides the shelter, creating a cold and warm ``aisle'' configuration, with the HVAC outlet located in the former.  The HVAC installation is capable of generating over 8000 CFM, and slight positive pressure in the cold reservoir serves to force air through hardware in the  rack, which is $> 90\%$ full.  (The static pressure created by the screened inlets is a nominal $0.03''$ H$_2$O.)

\section{Monitor \& Control}
\label{sec:gui}
\begin{figure*}
\begin{center}
\includegraphics[width=17cm]{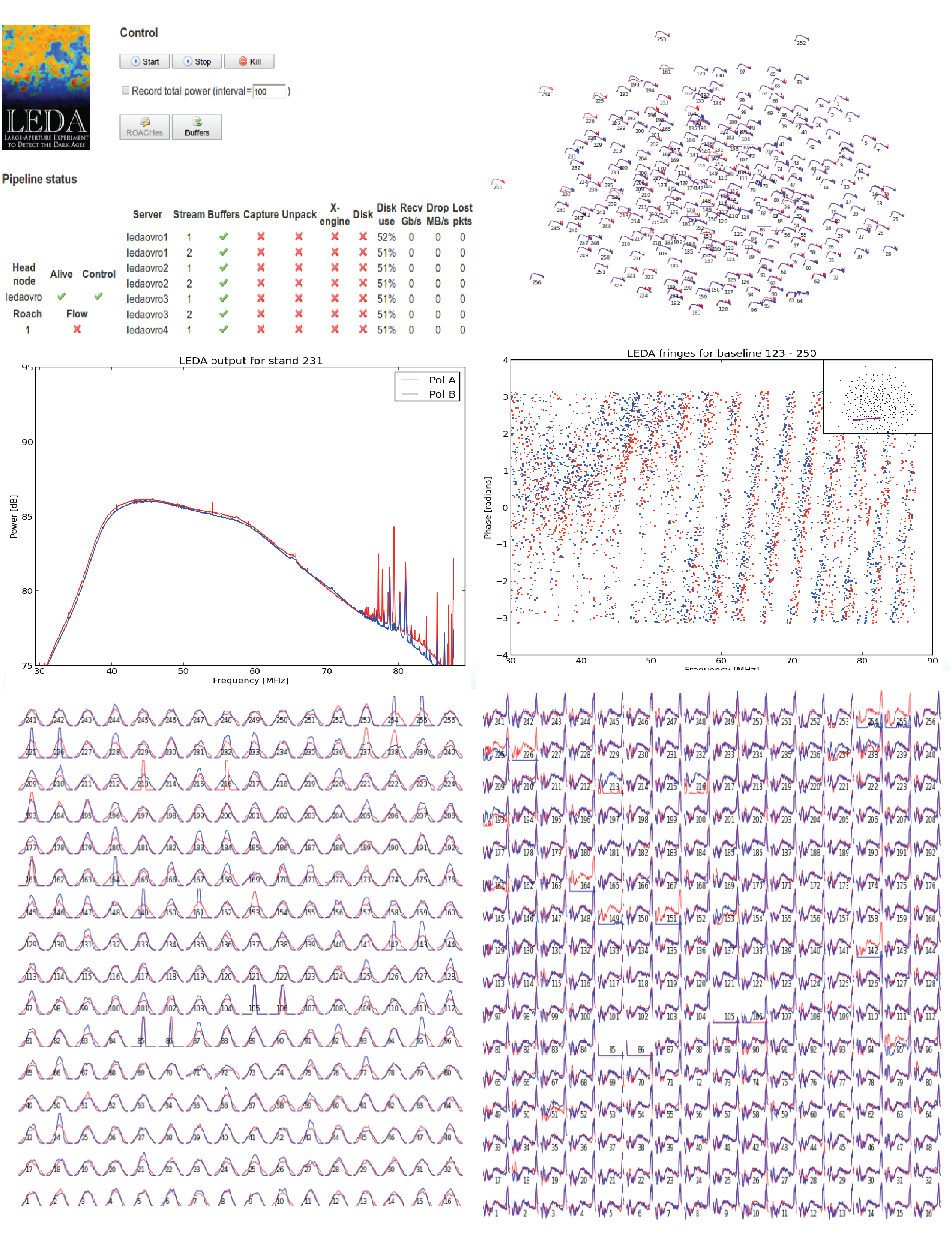}
\caption{M\&C display. ({\it Top left:}) The main control screen on which the status of the data processing pipelines for the correlator, and primary controls (start, stop, kill) are displayed. ({\it Top right:}) The bandpasses for each antenna, positioned as they are in the array though not to scale. ({\it Middle left:}) The correlator output for a single antenna (magnified) which may be selected interactively. ({\it Middle right:}) The visibility phase for a selected antenna pair. ({\it Bottom left:})  The histograms of raw ADC samples that highlight signal paths with anomalies and to which flagging must be applied. ({\it Bottom right:}) The spectra from the raw ADC samples at left.  Paths with red or blue only display are missing one polarization.  \label{fig:MandC}}
\end{center}
\end{figure*}

General control for the LEDA correlator requires only on and off commands, as the observations are topocentric and free running. However, data display that enables simultaneous monitoring of system status for such a large number of inputs and products poses a substantial challenge.  The correlator GUI is provided through a web interface (Figure \ref{fig:MandC}) that combines information from \texttt{PSRDADA} logs, and graphical renderings of configurations and streaming data on a single screen (e.g., side-by-side tabulation of packet loss for each antenna, real time monitoring of all autocorrelation spectra, and ADC time series and sample histograms), allowing the user to monitor all aspects of the system simultaneously.  As the correlator processing is split between the server nodes by frequency channel, the band must be reconstructed by taking a portion of data from each processing node. This is accomplished by a series of scripts on each server that read data from the most recent file written to the server local data storage and send this to the headnode, via sockets. Keeping data in separate streams (one for each GPU), somewhat counter-intuitively, is helpful in that it enables scaling of the display, e.g., to serve larger arrays, without increasing the per server load.

\section{First light}
\label{sec:first light}

\begin{figure}
\begin{center}
\includegraphics[width=9cm]{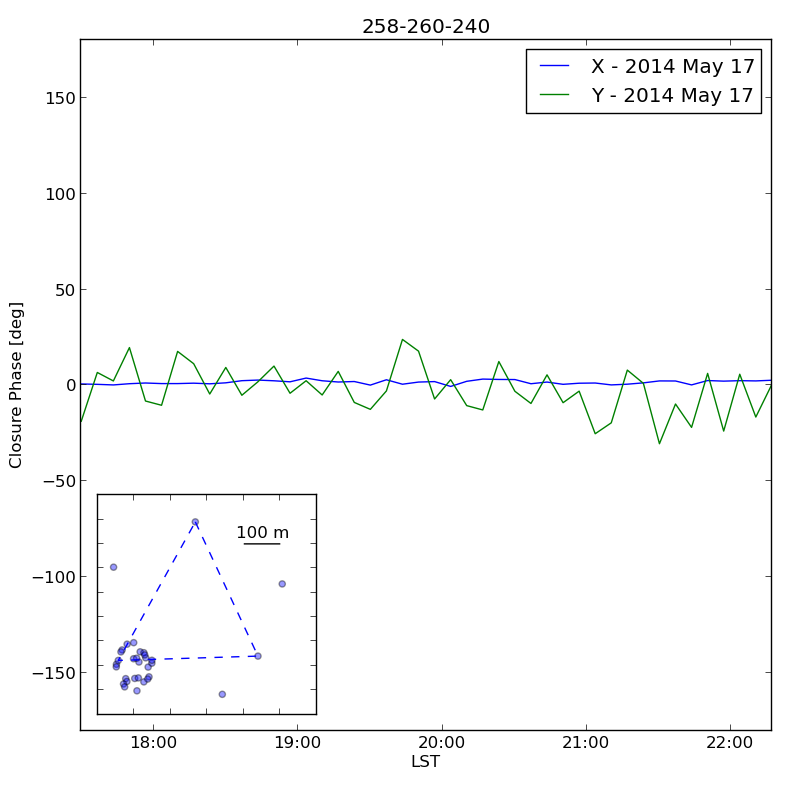}
\caption{Closure phase between LWA1 stands 258, 260 and 240.  Observations made over a 4~hr transit of Cyg\,A at 50\,MHz, 1\,MHz bandwidth. Each plot point is an average of 400\,seconds of data. The impact of RFI  on the RMS of the Y polarization residual is clearly visible.} \label{fig:closure}
\end{center}
\end{figure}

\begin{figure}
\begin{center}
\includegraphics[width=9cm]{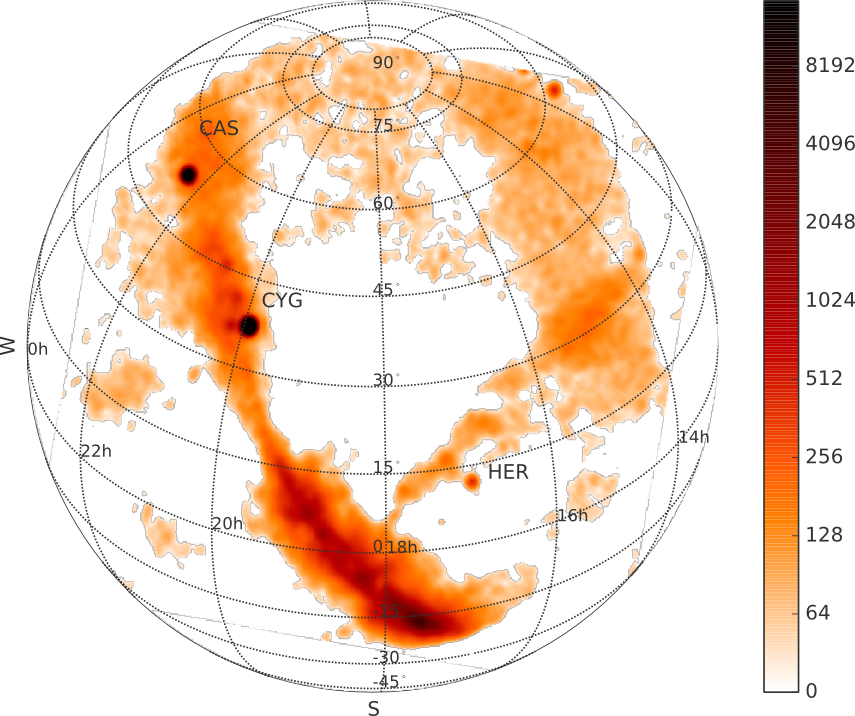}
\caption{Instantaneous all-sky dirty image from LWA-OV using the LEDA512 correlator at night on 2013 August 26 (2.616\,MHz centered at 47.004\,MHz and 2.13 seconds).  The flux scale is normalized to 10$^4$\,Jy for Cyg\,A. Dominant structures are Cyg\,A,  Cas\,A, Her\,A, and the Galactic plane.} 
\label{fig:image1}
\end{center}
\end{figure}

Verification tests for the correlator were performed both in lab and on sky.   The F and X-engines were tested individually: the fixed-precision FPGA firmware using simulated data and injection of continuous wave test tones into a fully assembled F-engine node; and direct comparison  of output from a CPU reference code  and the \texttt{xGPU} package, with agreement to floating point machine precision.  Initial sky tests are reported here, including estimates of closure phase, frequency measurement for a known source of narrowband interference,  and sky imaging.  At the LWA1 site, where a 64 input version of the correlator is deployed, measurements of closure phase were made for several hours on either side of transit for Cyg\,A. These measurements were generally consistent with the zero mean residual anticipated for signals dominated by a single point source (Figure\,\ref{fig:closure}), however the RMS of the one polarization (Y) residual is higher than expected for a point source based on the radiometry equation and for point source confusion (scaled from a 150\,MHz measurement by \citet{bernardi2013b}). This is not surprising, due to the large amount of polarized RFI in the band at these frequencies. Additional small variations from the predicted RMS are also expected, as the sky structure is not point-like, rather only dominated by Cyg\,A, and the data may also be affected by as yet uncalibrated ionospheric refraction. At both LWA1 and LWA-OVRO site, analogous to the injected signal during lab testing, intermittent RFI lines from amateur radio and digital TV were used to ensure the frequency axis was aligned correctly.  Figure\,\ref{fig:image1} shows the first-light sky image from the LWA-OVRO correlator, obtained from a 2.13 second snapshot on 2013 August 26, achieved with output from just a single X-engine GPU, i.e., 109 frequency channels or 2.616\,MHz bandwidth, centered at 47.004\,MHz.  The data were processed with the \texttt{CASA} package\cite{asclCASA} using Cyg\,A as a flat spectrum point source calibrator normalized to $10^4$\,Jy \citep[based on ][]{baars1977}.    The image compares well to that obtained by the LWA1 Prototype All-Sky Imager (PASI) and displayed via the LWA-TV system\footnote{See \url{http://www.phys.unm.edu/~lwa/lwatv.html}.} \cite{oben2014}, though the LWA-OV image is of higher angular resolution (1.8$^\circ$).

\section{Conclusions and Future Directions}

Initially scaled from 32 to 512 inputs, the implementation of the correlator presented could continue to scale to of order 5000 antennas. Scaling past this point would require modification of the X-engine code, currently capable of scaling only until the limit of one frequency channel per GPU is reached. The requirements for the LEDA correlator, while high in input, were comparatively low in bandwidth compared to the GHz level bandwidths available in other current, large,  systems. The modular design on the correlator allows for extensions to be made in this direction if required. Tradeoffs that need to be made include the number of inputs per ROACH2, and the number of servers required to process the data. Discussion of adapting the correlator to overcome these limitations and the required tradeoffs can be found in K14.

The correlator currently requires approximately 9~kW of power. Much of this, $\approx$3.6\,kW, is due to the servers required to host the GPUs. There are several future development avenues that will reduce this figure. Running on  the recently released Maxwell GPU architecture, the \texttt{xGPU} correlation code is expected to reach, if not exceed, the 80\% usage obtained on the Fermi GPU architecture \cite{clark2012}, immediately giving a 45\% increase in the processing available on each GPU. If the current correlator was replicated with the NVIDIA 980 GM204 architecture,  a reduction in power of $\approx$2.06~kW would be expected, due to the increased power and processing efficiency of the 980, and also the reduced number of servers required. Future developments will also explore the ability to use RDMA over converged ethernet (RoCE) and GPU-direct developments to reduce the server CPU requirements by bypassing the CPU for data transfer. In the current correlator, for example, replacing the current 115\,W CPUs with $\approx$10\,W ARM processors would save approximately 2.3\,kW of power compared to the current design. 

Deployed at the Owens Valley Radio Observatory,  the LEDA correlator, based on the K14 architecture, serviced the largest number of inputs of any correlator in operation at the time: 512, processing 57.552~MHz of bandwidth.  While a relatively small bandwidth, LEDA-512 has processing requirement approaching that of larger systems due to the high number of inputs. Taking the measure of correlator computational requirement presented in \citet{daddario2011}, where the size $S=N^2B$, where $N$ is the number of antennas and $B$ the bandwidth, LEDA-512 is measured at $3.77\times 10^{12}$~Hz, this is of similar ($\approx 60\%$) order to the correlator required for the VLA, assuming 8~GHz bandwidth, at $5.82\times 10^{12}$~Hz, and $\approx 11\%$ of the largest correlator developed to date, ALMA, at $3.28\times10^{13}$~Hz. Designed and implemented in 18-months, the correlator architecture demonstrates the benefits of hybrid systems, both in terms of development time, and simplicity.

\section*{Acknowledgments}

Research presented here was supported by National Science Foundation grants PHY-083057, AST-1106045, AST-1105949, AST-1106059, AST-1106054, and OIA-1125087

The authors acknowledge contribution from the Long Wavelength Array facility in New Mexico, which is supported by the University Radio Observatories program under grant AST-1139974, and National Science Foundation grant AST-1139963. The authors also thank Xilinx and NVIDIA for hardware and software contributions, Digicom Electronics for expedited manufacture and delivery of  production ROACH systems and ADCs early in the first production cycle, server vendor Silicon Mechanics for configuration testing, and Blundell, Tong, Leiker, and Kimberk of the Smithsonian Astrophysical Observatory Receiver Lab for expert discussion and construction of critical RF prototypes and field deployable hardware.

\clearpage


\begin{thebibliography}{9}


\bibitem[Baars et al.(1977)]{baars1977} Baars, J.W.M., Genzel, R., Pauliny-Toth, I.I.K., Witzel, A., 1977, A\&A, 61, 1, 99-106

\bibitem[Bernardi et al.(2013)]{bernardi2013b} Bernardi, G., Greenhill, L.~J., Mitchell, D.~A., et al.\ 2013, ApJ, 771, 105

\bibitem[Bernardi, McQuinn \& Greenhill(2014)]{bernardi2014} Bernardi, G., McQuinn, M., Greenhill, L. J., 2014, submitted ApJ, arXiv:1408.0411 

\bibitem[Bowman \& Rogers(2010)]{bowman2010} Bowman, J. D., Rogers, A. E. E. 2010, Nature, 468, 796-798

\bibitem[Clark, La Plante \& Greenhill(2012)]{clark2012} Clark, M. A., La Plante, P. C., Greenhill, L. J. 2012, IJHPCA, DOI 10.1177/1094342012444794

\bibitem[D'Addario (2011)]{daddario2011} D'Addario, L., 2011, SKA Memo 133

\bibitem[Edgar, et al.(2010)]{edgar10} Edgar, R. G., Clark, M. A., Dale, K., Mitchell, D. A., Ord., S. M., Wayth, R. B., Pfister, H., Greenhill, L. J. doi:10.1016/j.cpc.2010.06.019

\bibitem[Ellingson et al.(2013)]{ellingson2013} Ellingson, S. W., Craig, J., Dowell, J., Taylor, G. B., \& Helmboldt, J. 2013, in IEEE Symp. on Phased Array Systems \& Technology, pp. 776-783

\bibitem[Greenhill \& Bernardi(2011)]{greenhill2011} Greenhill, L. J. \& Bernardi, G. 2012, in 11th Asian-Pacific Regional IAU Meeting 2011, NARIT Conference Series vol. 1, eds. S. Komonjinda, Y. Kovalev, and D. Ruffolo, (Bangkok, NARIT)

\bibitem[Int. Cons. Sci.(2011)]{asclCASA} International Consortium Of Scientists, 2011, CASA, Astrophysics Source Code Library, 1107.013

\bibitem[Kocz et al.(2014)]{kocz2014} Kocz, et al. 2014, JAI, 3, DOI: 10.1142/S2251171714500020

\bibitem[Mitchell, et al.(2008)]{mitchell2008} Mitchell, et al. 2008, ISTSP, 2, 5, 707-717

\bibitem[Obenberger et al.(2014)]{oben2014} Obengerger, K.S., Taylor, G.B., Hartman, J.M., Clarke, T.E., Dowell, J., Dubois, A., Dubois, D., Henning, P.A., Lazio, J., Michalak, S., Schinzel, F.K., 2014, submitted JAI

\bibitem[Parsons, et al.(2006)]{parsons2006} {Parsons}, A., {Backer}, D., {Chang}, C., {Chapman}, D., {Chen}, H., {Crescini}, P., {de Jesus}, C., {Dick}, C., {Droz}, P.,  {MacMahon}, D., {Meder}, K., {Mock}, J., {Nagpal}, V., {Nikolic}, B., {Parsa}, A., {Richards}, B., {Siemion}, A., {Wawrzynek}, J., {Werthimer}, D., {Wright}, M., 2006, "PetaOp/Second FPGA Signal Processing for SETI and Radio Astronomy," Signals, Systems and Computers, 2006. ACSSC '06. Fortieth Asilomar Conference on , vol., no., pp.2031,2035, Oct. 29 2006-Nov. 1 2006 doi: 10.1109/ACSSC.2006.355123

\bibitem[Parsons, et al.(2008)]{parsons2008} {Parsons}, A., {Backer}, D., {Siemion}, A., {Chen}, H., {Werthimer}, D., {Droz}, P., {Filiba}, T., {Manley}, J., {McMahon}, P., {Parsa}, A., {MacMahon}, D., {Wright}, M., 2008, PASP, 120, 1207-1221

\bibitem[Parsons, et al.(2009)]{parsons2009} {Parsons}, A., 2009, Signal Processing Letters, IEEE, 16, 6, 477-480, DOI: 10.1109/LSP.2009.2016836

\bibitem[Thompson, Emerson \& Schwab(2007)]{thompson2007} Thompson, A.R., Emerson, D.T., Schwab, F.R., Radio Science, 2007, 42, DOI: 10.1029/2006RS003585

\bibitem[Pritchard \& Loeb(2010)]{pritchard2010} Pritchard J.R., Loeb A., 2010, Phys. Rev. D, 82, 3006

\bibitem[Taylor et al.(2012)]{taylor2012} Taylor, G.B., et al. 2012, JAI, 1, DOI: 10.1142/S2251171712500043

\bibitem[Thompson, Moran, \& Swenson(2001)]{tms01} Thompson, A. R., Moran, 
J. M., \& Swenson, G. W., Jr. 2001, Interferometry and Synthesis in Radio 
Astronomy (New York: Wiley-Interscience)

\end{thebibliography}
\end{document}